\input{aipcheck}

\documentclass[
    ,final            % use final for the camera ready runs
 ,numberedheadings % uncomment this option for numbered sections
  ]
  {aipproc}

\layoutstyle{6x9}

\begin{document}

\title{Soft-excess in ULX spectra: \\ 
disc emission or wind absorption?}


\author{A. C. Gon\c{c}alves}{
address={LUTH, Observatoire de Paris-Meudon, 5 Place Jules Janssen, 
F-92195 Meudon, France}, 
altaddress={CAAUL, Observat\'orio Astron\'omico 
de Lisboa, Tapada da Ajuda, P-1349-018 Lisboa, Portugal} 
}

\author{R. Soria}{
address={Harvard-Smithsonian Center 
for Astrophysics, 60 Garden st, Cambridge, MA 02138, 
USA}, 
altaddress={Mullard Space Science Laboratory (UCL), 
Holmbury St Mary, Dorking, Surrey, RH5 6NT, UK }
}

\begin{abstract}
We assess the claim that Ultra-luminous X-ray sources (ULXs) 
host intermediate-mass black holes (BH) by comparing the cool 
disc-blackbody model with a range of other models, namelly 
a more complex physical model based on a power-law 
component slightly modified at various energies by smeared 
emission/absorption lines from highly-ionized gas. 
Our main conclusion is that the presence 
of a soft excess, or a soft deficit, depends entirely 
on the energy range to which we choose to fit the ``true'' 
power-law continuum;  hence, we argue that 
those components should not be taken as evidence 
for accretion disc emission, nor used to infer BH masses. 
We speculate that bright ULXs could be in a spectral 
state similar to (or an extension of) the steep-power-law 
state of Galactic BH candidates, in which the disc is 
completely comptonized and not directly detectable, 
and the power-law emission may be modified 
by the surrounding, fast-moving, ionized gas. 
\end{abstract} 

\maketitle

%%%%%%%%%%%%%%%%%%%%%%%%%%%%%%%%%%%%%%%%%%%%
%% MAINMATTER
%%%%%%%%%%%%%%%%%%%%%%%%%%%%%%%%%%%%%%%%%%%%

\section{The nature of ULXs: stellar-mass binary systems, 
or intermediate-mass black holes?}
Ultraluminous X-ray sources (ULXs) are point-like, off-centre,
accreting X-ray systems, with apparent isotropic X-ray luminosities
up to a few $10^{40}$ erg s$^{-1}$, that is, they can be
almost two orders of magnitude more luminous than the Eddington
limit of typical Galactic black hole candidates (BHCs).
The main unsolved issue is whether ULXs are powered by black holes
(BHs) more massive than the BHCs, perhaps in the intermediate-mass
range ($\sim 10^{3}~M_{\odot}$; Miller et al.~2004), or
by stellar-mass BHs accreting at super-Eddington rates (Begelman 2002);
alternatively, their brightness could be due to beaming
along the line-of-sight of the observer
(e.g. King et al.~2001).

The standard way to determine the mass 
of an accreting black hole in X-ray binaries is based 
on phase-resolved spectroscopic and photometric studies 
of their optical counterparts. Attempts to apply similar 
techniques to ULXs have been  
inconclusive  so far, mostly because  of their optical faintness. 
%At typical distances of a few Mpc, 
%($V \sim 24$ for most optical counterparts).  
In many cases, crowding is also a problem: 
the X-ray error circle may be consistent with 
an unresolved group of stars. Pioneering efforts 
(e.g., Gris\'{e} et al. 2006) 
may yield results for a few sources in the near future. 
Meanwhile, though, one has to rely on indirect methods 
to estimate the BH mass:  either X-ray timing analysis  
(e.g. breaks in the Power Density Spectrum; Quasi-Periodic 
Oscillations), or spectral studies (e.g. X-ray data from 
{\it ASCA}, {\it Chandra} and {\it XMM-Newton}).

\section{The X-ray spectra of ULXs}
\subsection{Black hole mass determination}
One way of determining the black hole mass of ULXs 
is based on X-ray spectral fitting over the 
``standard'' $0.3$--$10$ keV band. In Galactic BHCs, 
the X-ray spectrum consists of essentially two 
components (power-law and thermal) with varying 
normalizations and relative contributions in various 
spectral states. The power-law component is scale-free 
and without a direct dependence on BH mass; however, 
its slope and normalization are related to the spectral 
state and normalized luminosity, the slope being flatter 
in the low/hard state. In what concerns  the thermal 
component, this is interpreted as the spectrum of an 
optically-thick Shakura-Sunyaev disc and contains, 
in principle, a direct dependence on disc size and BH mass. 
It is possible to provide a reasonable estimate 
of the BH mass in Galactic BHCs by modeling such a thermal 
component in XSPEC, through more or less complex 
implementations of the disc-blackbody model.  
%(e.g.  {\tt diskbb}, {\tt diskpn}).   
It seems reasonable to apply the same simple tools 
to estimate the mass of the accreting BHs in ULXs, 
if they are scaled-up versions of Galactic BHCs.

\subsection{Hot-disc vs. cold-disc models} 
For about a dozen of the brightest ULXs, it was noted 
(e.g. Miller et al.  2004) 
that the $0.3$--$10$ keV spectrum is dominated by  
a featureless broad-band component, interpreted as a power-law 
plus a ``soft-excess'' significantly detected below 1 keV;  
for various sources, an additional thermal component 
with $kT \sim 0.1$--$0.2$ keV seems to lead to better fits. 
We call this interpretation the ``cool disc'' (CD) model. 
By analogy with Galactic BHCs, one can attribute true physical
meaning to such phenomenological fits, interpreting the fitted
temperature as the color temperature near the inner boundary
of an accretion disc; this approach  
does not assume or require a specific physical 
model for the power-law component.
If we apply the standard disc relation\footnote{In fact, this relation
may not be applicable to power-law dominated ULXs. We suggest
elsewhere in these Proceedings (Soria et al.~2006) that in the
framework of the CD scenario, a more accurate relation leads
to a mass estimate $\sim 50~M_{\odot}$. However here we stick
to a discussion of the ``convential'' interpretation
of the CD model as a signature of intermediate-mass BHs (IMBHs).}  
between temperature, luminosity and BH mass, 
we obtain characteristic mass values $\sim 10^{3}~M_{\odot}$. 

However, the CD model fitting is not unique.
Good fits are also obtained, for many ULXs, using a disc
with color temperature $\sim 1.5$--$2.5$~keV, which contributes
most of the flux above 1 keV, plus a soft excess
with characteristic blackbody temperature $\sim 0.2$~keV.
We refer to this scenario as the ``hot disc'' (HD) model
(Stobbart et al.~2006). A possible physical interpretation
of the phenomenological HD model is that the harder emission
comes from a ``slim disc'' (e.g. Watarai et al.~2001), 
% Watarai et al.~2001; Ohsuga et al.~2005; Heinzeller et al.~2006 
typical of highly super-Eddington mass inflow rates 
onto stellar-mass BHs, with significant photon trapping. 
The softer component could be interpreted as 
downscattered photons in a cooler outflow or photosphere. 
An alternative approach, also successfully applied to 
Galactic BHCs, is to fit the X-ray spectra with a more 
complex, self-consistent model  
in which a power-law-like component arises as comptonized 
emission from seed thermal photons, upscattered in a corona. 
When such physical models are applied to bright ULXs, 
it is found (Goad et al. 2006; Stobbart et al. 2006) 
that the emission from the inner disc may be almost 
completely comptonized in a warm, optically-thick, 
but patchy corona.

\begin{figure}
  \centering
  \includegraphics[width=9cm]{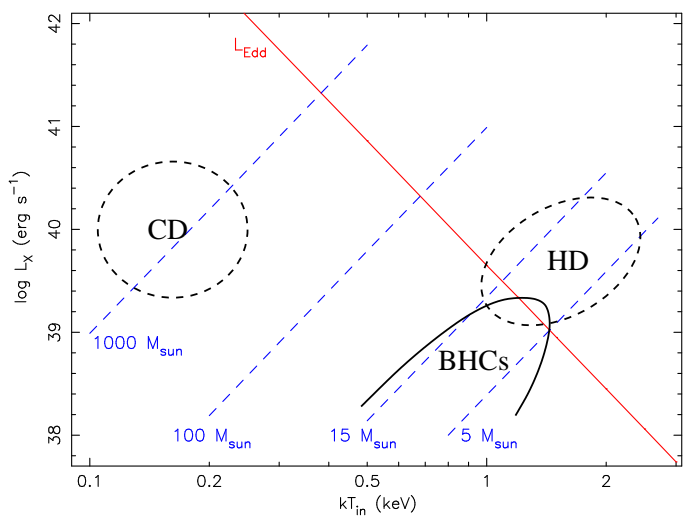}
      \caption{Location of 
Galactic BHCs and ULXs in a disc temperature vs.  
X-ray luminosity plot; adapted from figure 3 in Stobbart 
et al.~(2006). The CD (cold-disc) model implies that 
ULXs are intermediate-mass BHs, emitting well below 
their Eddington limit. The HD (hot-disc) model 
suggests that ULXs are stellar-mass objects (an extension 
of the Galactic BHC class), emitting above Eddington. }
       \label{ULXs}
   \end{figure}

CD and HD model make different assumptions and have been used to infer
different BH masses and physical structures of the accretion flow. 
BH masses $\sim 10^{3}\,M_{\odot}$, suggested by the CD model, require
more complicated and so far untested formation scenarios; furthermore, 
in the CD model, ULX luminosities remain always an order 
of magnitude below their Eddington limit (Fig.~1), 
even for the brightest sources, suggesting perhaps 
some kind of upper limit to the mass supply. 
This behaviour is not observed in Galactic  BHCs, which often 
reach, or even slightly exceed, their Eddington limit. 
On the other hand, the HD model suggests that 
ULXs could be stellar-mass BHs emitting 
at up to an order of magnitude above their 
Eddington limit. From a physical point of view, 
the two scenarios (stellar-mass but well above Eddington, 
or $\sim 10^3 M_{\odot}$ but well below Eddington) 
have different kinds of drawbacks, 
and more constraining observations in other 
energy bands will be necessary to rule either one out. 
%In the CD model, the disc directly contributes only $\sim 5$--$20\%$
%of the X-ray luminosity; instead, in the HD scenario, the disc
%emission is the dominant component of the X-ray spectrum.
In its simplest, phenomenological form, the CD model
does not explain the observed break in the spectral slope
at energies $\ge 5$ keV; that can be taken into account
only by more complex models in which the power-law
component is replaced by comptonized emission emerging
from a low-temperature corona. On the other hand, the HD
model overpredicts that break. Spectral observations
in the $\sim 15$--$30$~keV range  
will be crucial to determine whether that spectral feature
is a change in slope or an exponential cutoff, and will
provide a test between the two models.  
Finally, the inner-disc temperature in the CD model 
is in the same range as the characteristic temperature 
of the soft excess in Seyfert 1s, despite the large 
mass difference between ULXs and Active Galactic Nuclei (AGN). 
Both kinds of spectra can formally be well fitted with a cool 
disc-blackbody component plus a power-law, but more likely 
the soft excess in AGN could be explained 
by a combination of blurred emission/absorption lines and reflection 
(Gierli\'nski \& Done 2004; Chevallier et al.~2006); this could 
suggest that ULXs and stellar-mass BHs are completely 
separate classes of sources. Instead, characteristic 
disc temperatures in the HD model fall within, or close to, 
the range of stellar-mass BH temperature; this could suggest 
that ULXs are simply an extension of the stellar-mass class 
at higher accretion rates.

\subsection{Disc emission or wind absorption?}
Both the phenomenological CD and HD models 
share the same bias: namely, that the dominant 
component of the spectrum is well determined  
by the observed emission at $\sim 2$--$5$~keV. For example, 
in the CD model, the spectrum is more or less 
a true-power-law in that energy range.
Deviations from the assumed true-power-law 
at energies $\le 2$~keV are thus cast in the form 
of a soft excess, while deviations at energies 
$\ge 5$~keV can be dismissed as small-count statistics, 
or with the introduction of an {\it ad hoc} cut-off, 
or by assuming a low-temperature corona. 
Similarly, the HD model assumes that the spectrum 
is a true disc-blackbody in that range, with its 
emission peak falling just below or around $5$ keV.
Again, this choice inevitably leads us 
to finding a soft excess below $2$ keV, modelled 
with an additional thermal component.
%A possible reason for this bias is that the continuum 
%in the $2$--$5$ keV spectral range is free from 
%line-of-sight cold or warm absorption, hardly modified 
%by any residual soft thermal-plasma emission, 
%and has good spectral resolution and sensitivity 
%in {\it Chandra} and {\it XMM-Newton}. 
%Thus, if we use a power-law model, it appears natural 
%to adjust its slope to fit that energy range, 
%and then take care of any deviations.

Evidence for a change in the spectral slope 
in the $2$--$10$~keV band is given by Stobbart et al. (2006), 
who show that a broken power-law fit provides an improvement 
over a single power-law fit in 8 out of 13 ULXs; 
this supports the idea that most sources cannot be 
described by a single power-law continuum across the whole 
band (a similar degeneracy is in fact common to many ULXs). 
Thus, instead of estimating the continuum in the 
$2$--$5$~keV range, we could equally well assume that
the continuum in the region $\sim 5$--$10$~keV 
is the true expression of the power-law. 
If we do that, we find that most bright ULXs
have a distinctive ``soft deficit''.
We would then try to devise complex physical models 
to explain that deficit, or, more simply, 
we would use phenomenological 
models. By analogy with the CD model, where  
a disc-blackbody component is used to account 
for the smooth, broad-band soft excess, 
we could select a smooth, broad-band 
absorption component. Figure 2 shows that both 
the soft deficit (left-hand panel) and 
the soft excess model (middle panel)  account 
equally well for the observations.

\begin{figure}
  \centering
  \includegraphics[width=13.5cm]{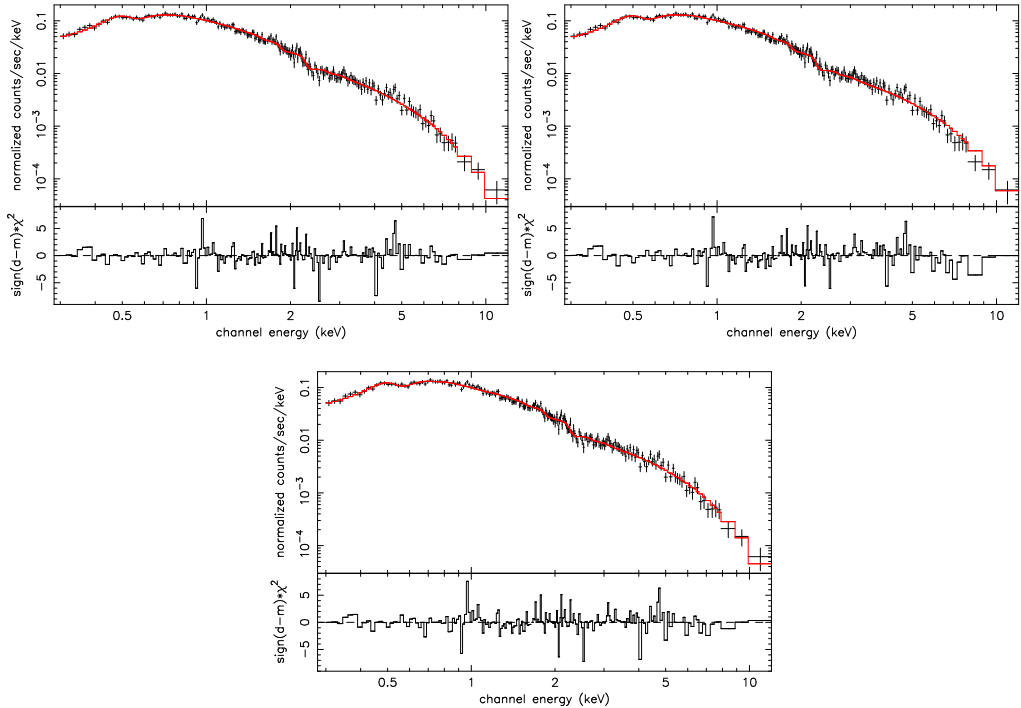}
      \caption{Three statistically-good fits 
to the {\it XMM-Newton}/EPIC spectrum 
of NGC\,4559 X-1 with 3 different models. Top left-hand panel:
the spectrum is modelled with a steeper 
($\Gamma \sim 2.7$) power-law with a broad 
absorption feature approximated by a (negative) 
disc-blackbody component at $0.42$ keV. 
Top right-hand panel: 
the spectrum is modelled with an underlying flatter 
power-law ($\Gamma \sim 2.2$) plus a soft excess, 
approximated by a (positive) disc-blackbody component 
at $0.14$ keV. We argue that 
neither the positive nor the negative disc-blackbody  
component has any physical meaning or relation with 
the accretion disc; they are simply convenient, 
versatile components to model broad bumps. 
Lower central panel: 
the same spectrum, modelled with an underlying power-law 
($\Gamma \sim 2.7$) modified self-consistently 
by smeared emission and  
absorption lines caused by a layer of highly 
ionized gas. The best-fit parameters of these  
three models, and other pertinent information, are 
given in Gon\c{c}alves \& Soria (2006).}
       \label{NG4559}
   \end{figure}

Less phenomenological, but more complex physical models, 
would of course show that such a soft deficit is not due to 
a negative disc-blackbody spectrum, but for example to 
smeared absorption lines.  We have shown one 
recent implementation of such complex models, 
which we have developed thanks to the photoionization 
code  TITAN, imported into XSPEC,  
and applied to two bright ULXs as an illustrative example 
(Gon\c{c}alves \& Soria 2006). Our modelling, illustrated 
in Fig.~2 (right-hand panel) with the spectrum of NGC~4559 X-1, 
shows that it is possible to produce broad, 
smooth emission, and absorption features, when 
an injected power-law  spectrum is seen through a 
highly-ionized plasma with midly relativistic motion.  
This hypothesis seems  less problematic than attributing 
those features to accretion disc emission.

\section{Conclusions}
The uncritical use of temperature-mass relations from the
disc-blackbody model has led to claims of BH masses
$\sim 10^{3}~M_{\odot}$, skewing both observational and theoretical
studies of ULXs towards the IMBH scenario.
We point out that this interpretation is far from unique.
Specifically, we argue that: {\it (i)} the CD model may be correct,
but the BH mass may be only $\sim 50 M_{\odot}$ (see our companion
paper in these proceedings); {\it (ii)} for many ULXs, both the CD and
HD models may be consistent with the observations; the latter
scenario corresponds to a super-Eddington stellar-mass BH;
{\it (iii)} both the CD and HD models may be incorrect: the deviations 
from the power-law spectrum could be due to reprocessing
in an ionized outflow. This radically alternative ULX scenario
is analogous to models previously applied to AGN. It implies
that the deviations from a pure power-law spectrum are not
related to disc emission, and therefore tell us nothing
about the BH mass. Without this piece of information,
the remaining evidence in favour of IMBHs is much weakened
for the majority of ULXs.
Both the CD and the ionized outflow models imply that
the X-ray spectrum is strongly dominated by a power-law
component, with the disc emission being either comparatively
small or entirely negligible. 

A somewhat similar situation
appears to occur in the steep-power-law state of Galactic BHCs.
Goad et al.~(2006) suggested that the temporal variability
of Holmberg II X-1 is similar to that found in the Galactic 
BHC GRS~J1915$+$105 ($M_{\rm BH} \approx 15 M_{\odot}$) in its
steep-power-law state.  
We speculate that some ULXs represent a further spectral 
state, contiguous to the steep-power-law state, in which 
the disc contribution is entirely negligible 
and, in addition, the dominant power-law component 
is modified by smeared emission and absorption 
from the surrounding, highly-ionized, possibly 
outflowing gas. Interestingly, one of the effects 
of the broad absorption features at $\sim 1$ keV 
is to make the continuum appear flatter 
than the injected power-law, 
over the $2$--$10$ keV range, as we noted 
when comparing positive and negative disc-blackbody 
models (cf. Fig.~2). 
This may be one reason why many bright ULXs in this class 
appear to have a flatter power-law slope when fitted
with a CD model, than Galactic BHCs in the steep-power-law 
state (the latter presumably being less affected 
by highly-ionized, fast outflowing plasma). 

Such a spectral state could be shared by higher-mass 
accretors such as AGN. Narrow Line Seyfert 1s, in particular, 
display a soft X-ray excess and characteristic variability which 
could be associated with a steep-power-law state. 
It has been shown (Chevallier et al.~2006) that the soft 
excess in AGN could be fitted with the same relativistically 
smeared ionized plasma model applied here to ULXs. Thus, our approach 
offers a possible common explanation to the properties 
of ULXs, soft-excess AGN and Galactic BHs; 
it suggests that the main spectral features in this 
bright state depend on the physical parameters of 
the outflowing plasma, not on the mass of the accretor.

%%%%%%%%%%%%%%%%%%%%%%%%%%%%%%%%%%%%%%%%%%%%%%%%

\begin{theacknowledgments}
%We thank Zdenka Kuncnic and Mat Page for fruitful discussions. 
ACG acknowledges support from the 
{\it Funda\c{c}\~ao para a Ci\^encia e a Tecnologia (FCT)},  
Portugal, under grant 
BPD/11641/2002. RS acknowledges support from an OIF 
Marie Curie Fellowship, through University College London.
\end{theacknowledgments}



\bibliographystyle{aipprocl} % if natbib is missing

%\endinput

%\bibliography{sample}

%  \typeout{******************************************}
%  \typeout{** Please run "bibtex \jobname" to optain}
%  \typeout{** the bibliography and then re-run LaTeX}
%  \typeout{** twice to fix the references!}
%  \typeout{******************************************}

\end{document}